# Beam Dump Window Design for CSNS

NIE Xiao-Jun(聂小军)[1]　LIU Lei(刘磊)[1]　WANG Hai-Jing (王海静)[1,2]

Institute of High Energy Physics, Chinese Academy of Sciences, Beijing 100049, China

[1] Institute of High Energy Physics, Chinese Academy of Sciences, Beijing 100049, China

[2] University of Chinese Academy of Sciences, Beijing 100049, China

**Abstract:** Beam dump window is one of the most important parts of beam dump of China Spallation Neutron Source (CSNS). The material, structure and other concerned issue have been discussed in this paper. Thermal analyses have been applied for the beam dump window design. By comparison, Glidcop®AL-15 and 316L have been chosen as window material. The window section has been designed as inner convex spherical surface and the window thickness has been set 1.5mm to 3mm by structure optimization. The window safety has been confirmed under the damage of magnet. All these analyses proved that the window can meet the requirements of CSNS beam dump well.

## 1 Introduction

Beam dump is one of the important devices in CSNS, which is used to prevent and shield the deserted beam. Four beam dumps named LRDUMP, LRDUMP1, INDUMP and RCSDUMP respectively have been designed for CSNS. The beam parameter of beam dump is listed as table 1 [1].

Table 1 Beam parameter for beam dumps

| Beam dump | Power (kW) | Energy (MeV) | Beam Size (X/Y, mm) |
|---|---|---|---|
| LRDUMP | 4.0 | 80 | （122,67） |
| LRDUMP1 | 0.2 | 80 | （19,26） |
| INDUMP | 2.0 | 80 | （51,22） |
| RCSDUMP | 7.5 | 1600 | （89,89） |

Beam dump window is one of the key devices of beam dump. It is located at the end of vacuum box, which is used to separate the shielding area and high vacuum region. The high vacuum needs to be kept when the beam penetrates the window. The window should have good thermodynamic characteristic and mechanical strength. Many factors, such as the window structure, material and energy deposition, need to be considered comprehensively during the beam dump window design.

## 2 Analytical method and energy deposition calculation

The temperature and stress of beam dump window are just simulated by ANSYS during the design. The steady and transient analyses are applied to confirm the window characteristics. Steady-state analysis shows the final temperature and stress of the window. And the transient analysis reveals the trend of temperature by the time [2].

The energy deposition is calculated using SRIM. The beam distribution shows as 2D Gaussian distribution. The beam energy loss and its distribution are calculated by Eq. (1) and Eq. (2) respectively.

$$Q = P/E \times dE/dx \times t. \quad (1)$$

1) Email: niexj@ihep.ac.cn

$$q(x, y) = \frac{Q}{2\pi\sigma_x\sigma_y} \times \exp^{-(\frac{x^2}{2\sigma_x^2}+\frac{y^2}{2\sigma_y^2})}. \quad (2)$$

Q is the total beam energy deposition. P and E are beam power and energy respectively. $dE/dx$ is the power loss in each millimeter thickness. $q(x, y)$ is the beam distribution function.

## 3 Material selection

The window will generate much heat and suffer much radiation when the beam penetrates it. So it demands the material has excellent thermodynamic characteristic, high ultimate strength, low energy deposition and good anti-radiation performance. Meanwhile, the processing and welding properties should be good for the window small thickness [3-4].

Generally, the material with lower density will get less energy deposition in the same thickness. And the material with good thermal performance will get low temperature at the center. The material choice needs to consider all the related properties.

Currently, ISIS and SNS choose Inconel alloy as the window material, while J-PARC and ESS use aluminum alloy instead [4]. As to CSNS beam dump window, five types of material have been considered for comparison. They are inconel718, GlidCop®AL-15, 316L, A5083 and Ti-6Al-4V. The window structure keeps the same when comparing the material. The energy deposition, temperature and stress for different materials are listed as table 2 ~ table 4 respectively.

Table 2  Energy deposition for different materials (Unit: W)

| Beam Dump | Window thickness | Inconel718 | GlidCop® AL-15 | 316L | A5083 | Ti-6Al-4V |
| --- | --- | --- | --- | --- | --- | --- |
| LRDUMP | 2mm | 550.233 | 556.149 | 519.794 | 198.975 | 299.196 |
| LRDUMP1 | 2mm | 27.512 | 27.807 | 25.99 | 9.949 | 14.96 |
| INDUMP | 2mm | 275.116 | 278.075 | 259.897 | 99.488 | 149.598 |
| RCSDUMP | 2mm | 11.657 | 11.826 | 11.007 | 4.144 | 6.322 |

Table 3  Highest temperature for different materials (℃)

| Beam dump | Inconel 718 | GlidCop® AL-15 | 316L | A5083 | Ti-6Al-4V |
| --- | --- | --- | --- | --- | --- |
| LRDUMP | 5244 | 830 | 2891 | 453 | 3092 |
| LRDUMP1 | 1070 | 101 | 729 | 78 | 970 |
| INDUMP | 5058 | 479 | 3481 | 305 | 4425 |
| RCSDUMP | 140 | 44 | 110 | 38 | 116 |
| Melting point | 1260-1336 | 1082 | 1370-1400 | 590-638 | 1604-1660 |

Table 4  Maximal stress for different materials (MPa)

| Beam dump | Inconel 718 | GlidCop® AL-15 | 316L | A5083 | Ti-6Al-4V |
| --- | --- | --- | --- | --- | --- |
| LRDUMP | / | 193 | / | 225 | / |
| LRDUMP1 | / | 37.2 | 895 | 20.9 | / |
| INDUMP | / | 245 | / | 150 | / |
| RCSDUMP | 41 | 12.2 | 31.8 | 8.46 | 16.9 |
| Allowable stress (MPa) | 1100 | 255-300 | 290 | 190 | 880 |

The result indicates the window temperature and stress characteristics under different materials. The window material can be decided based on the tables. LRDUMP and

INDUMP only use GlidCop® AL-15 because the other materials have high temperature. LRDUMP1 can use GlidCop® AL-15 or A5083 and RCSDUMP can use any of the five materials. Considering the material purchase, manufacturing cost and the property of resistance to corrosion, Glidcop® AL-15 and 316L have been chosen as beam dump window materials finally. The charateristic shows as table 5 [5-6].

Table 5　Characteristic of two different window materials

| Material | Density (kg/m^3) | thermal conductivity (W/m-K) | thermal expansion coefficient (μm/m-°C) | Specific heat of fusion (J/kg-°C) |
|---|---|---|---|---|
| Glidcop®AL-15 | 8810 | 365 | 18.5 | 385 |
| 316L | 8000 | 16.3 | 16.00 | 500 |

| Material | yield strength (MPa) | modulus of elasticity (GPa) | Poisson's ratio | melting point (℃) |
|---|---|---|---|---|
| Glidcop®AL-15 | 255 | 130 | 0.343 | 1082 |
| 316L | 290 | 193 | 0.29 | 1370 |

## 4 Structure design

### 4.1 Window section shape optimization

For different beam power and material, three types of section shape can be considered: Single-layered, double-layered and multi-pipe structure. The single-layered window has simple structure, but the cooling condition is worse than the others. It can be used under low beam power. The double-layered and multi-pipe windows have better cooling condition and can be used under higher beam power. But the structure is more complex [3-4].

CSNS beam dump power is just between 0.2kw to 7.5kw. So the single-layered window is considered. And for the space restriction, it is difficult to set cooling device for beam dump window. So the window is just under natural cooling. Three single-layered window section shapes have been considered for contrast. They are flat, inner convex spherical surface and out convex spherical surface. They are shown as Fig.1. Here just use LRDUMP window as an example.

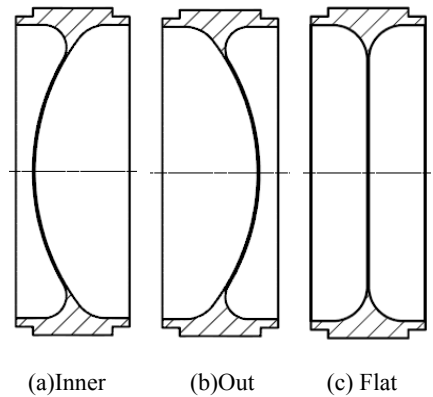

(a)Inner　　(b)Out　　(c) Flat

Fig.1 Three window section shapes

The temperature and stress for different windows can be simulated by ANSYS. The result is shown as table 6 and table 7.

Table 6　Window highest temperature (℃)

| Window structure | LRDUMP | LRDUMP1 | INDUMP | RCSDUMP |
|---|---|---|---|---|
| Sphere in | 416.3 | 101.1 | 283.5 | 144.6 |
| Sphere out | 419.3 | 101.4 | 279.6 | 143.1 |
| Flat | 401.2 | 101.1 | 278.7 | 141.3 |

Table 7　Window Maximal Stress (MPa)

| Window shape | LRDUMP | LRDUMP1 | INDUMP | RCSDUMP |
|---|---|---|---|---|
| Sphere in | 20.9 | 104 | 163 | 108 |
| Sphere out | 26.4 | 118 | 178 | 120 |
| Flat | 48.2 | 619 | 282 | 271 |

According to the result, it can be concluded that the temperature is almost the same for different window sections, but the stress has significant difference. Compared to the others, window with inner convex spherical surface has the best stress condition. window sections are Considering the beam power, energy and distribution of table 1, the beam dump window section shapes have been designed as Fig.2.

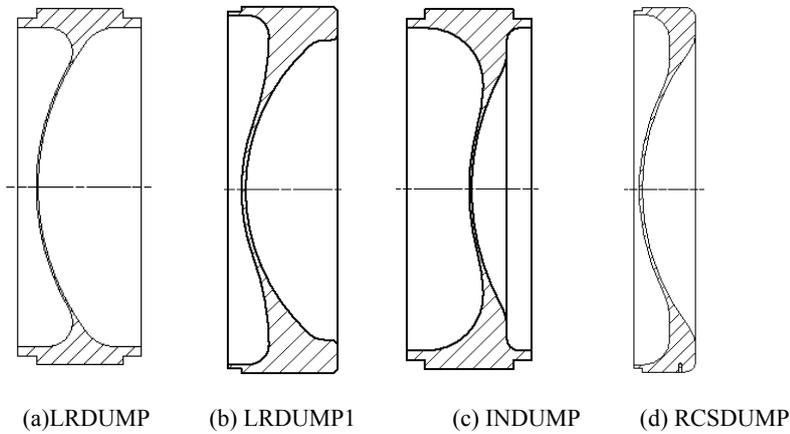

(a)LRDUMP   (b) LRDUMP1   (c) INDUMP   (d) RCSDUMP

Fig.2  Beam dump window section shapes

**4.2 Window thickness optimization**

Window thickness has great effect on the window characteristics. The thinner window can get less energy deposition, but the processing will become worse with weak mechanical strength. Increasing the window thickness can raise the mechanical strength and improve the processing property, but the energy deposition will increase accordingly. So it is necessary to make a balance between the thickness and the energy deposition.

Commonly, beam dump window thickness is about 1 millimeter to several millimeter, some even thinner than 1 millimeter. For example, SNS thickness is 2mm, while PEFP is just 0.5mm [7-8]. In order to get the reasonable thickness of CSNS beam dump window, the temperature and stress under different window thickness have been simulated by ANSYS. The trends of temperature and stress are shown as Fig.3.

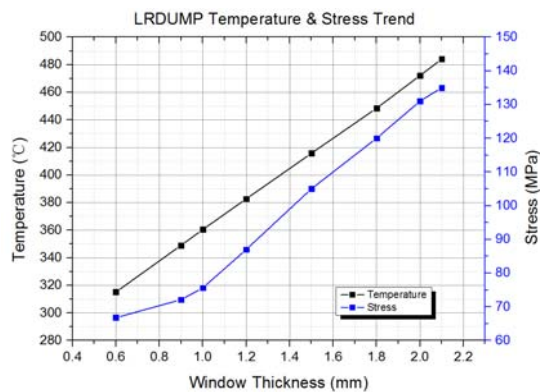
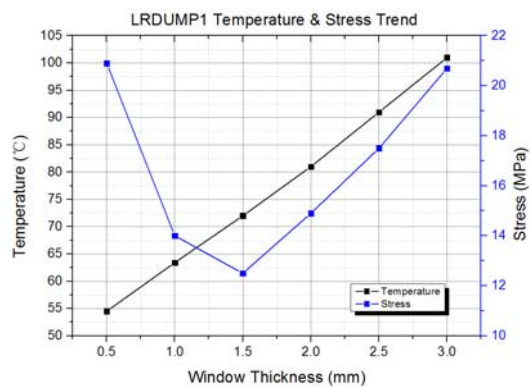

(a) LRDUMP    (b) LRDUMP1

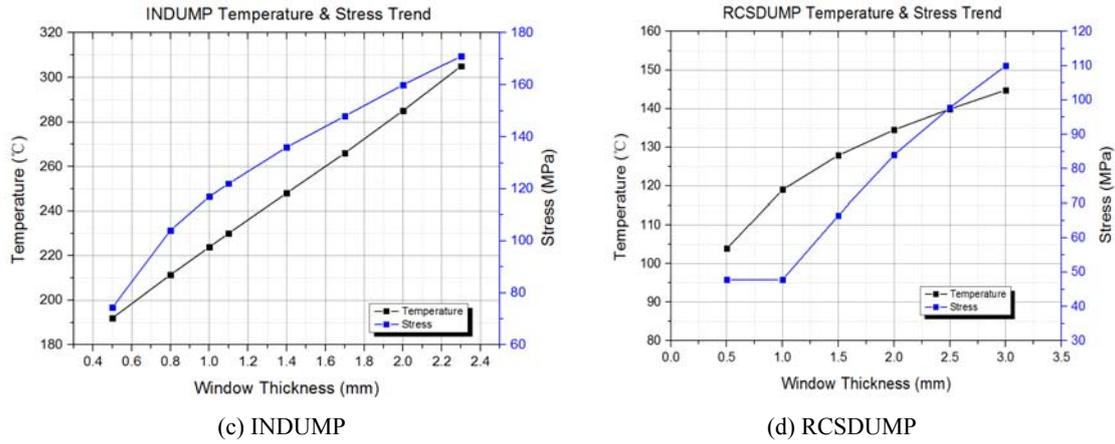

(c) INDUMP  (d) RCSDUMP

Fig.3 Beam dump window temperature & stress trend by thickness

The trend indicated that the max temperature and stress are increased by the window thickness. Thus the thickness of window should be as thin as possible. Based on Fig.3 with the consideration of processing technology, the thickness has been decided as table 8.

Table 8 Beam dump window thickness for CSNS (Unit: mm)

| Beam dump | LRDUMP | LRDUMP1 | INDUMP | RCSDUMP |
|---|---|---|---|---|
| Thickness | 1.5 | 3 | 2 | 3 |

## 5 Safety confirmation

Normally, the beam size is under the magnet control to meet the beam dump window requirement. The beam dump window can work safely under the normal beam size. But if the magnet becomes failing, the beam size will change significantly. The beam size on the beam dump window can be reached by the simulation of physics. They are listed as table 9.

Table 9 Beam size on beam dump window (Normal vs. failing)

| Beam Dump | LRDUMP | LRDUMP1 | INDUMP | RCSDUMP |
|---|---|---|---|---|
| Normal (X-mm/Y-mm) | 122/67 | 19/26 | 51/22 | 89/89 |
| Failing (X-mm/Y-mm) | 24/9 | 19/26 | No beam | 163/88 |

Note: The beam size indicates the range of beam action in two directions.

It is obvious that the failure of magnet has big effect on LRDUMP window while it has less effect on others. So the following analysis is just focused on LRDUMP window.

The beam size of LRDUMP will become small sharply when the magnet fails. So the window temperature and stress distribution will change accordingly. Fig.4 showed the steady-state temperature and stress under failing magnet. The maximal temperature and stress of the window has increased greatly. So the beam must be cut down as soon as possible under this situation.

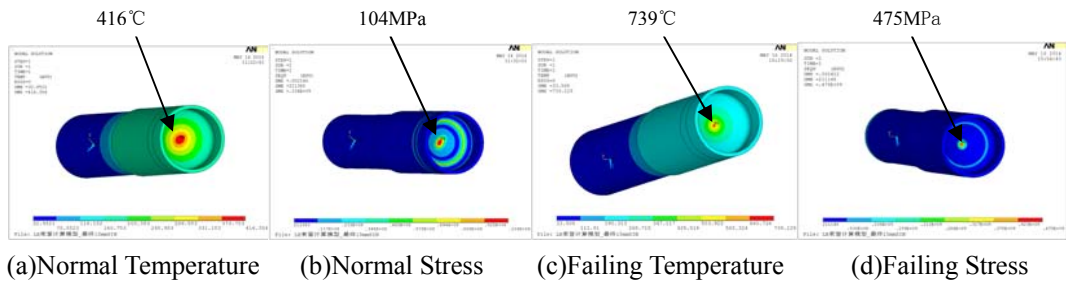

416℃    104MPa    739℃    475MPa

(a)Normal Temperature  (b)Normal Stress  (c)Failing Temperature  (d)Failing Stress

Fig.4 Temperature and stress under normal and failing magnet

It has been confirmed that the control system can make response within 40ms. Considering the action time, 1s should be enough for the beam stop. The window maximal temperature trend has been simulated by ANSYS transient state analysis. They are shown as Fig.5.

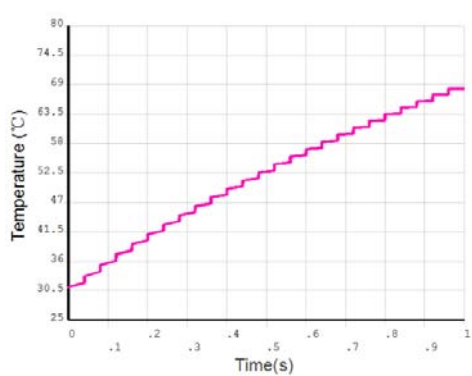
(a)Normal, 1s

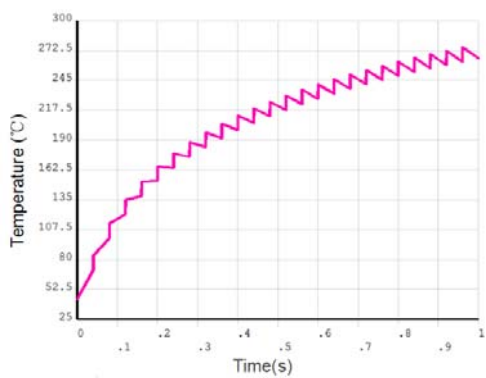
(b) Failing, 1s

Fig. 5 Window temperature curve under normal and failing magnet

It is obvious that the temperature increases slowly under normal magnet while it increases fast under failing magnet. But even under failing magnet, the temperature is not so high at the end of 1s. It is just about 270℃。So the window can be safe even if the magnet becomes broken.

## 6 Conclusions

Beam dump window is crucial to the beam dump. Its design includes material choice, structure design and other concerning issue. Glidcop®AL-15 and 316L have been chosen as window material for good thermodynamic characteristic. By comparison, the window section has been designed as single-layered section with inner convex spherical surface. Window thickness has been set between 1.5mm and 3mm by the window temperature and stress trend. Meanwhile, the working status shows that the window can be safe even if the magnet becomes broken. In summary, the property of the windows can meet the demands in theory.

colleagues for their support, help and advice.

## Acknowledgments

*The authors would like to thank their CSNS*

# CSNS 废束站束窗设计

聂小军 [1]，刘磊 [1]，王海静 [1,2]

1、中国科学院高能物理研究所，北京，10049

2、中国科学院大学，北京，10049

**摘要：** 废束站束窗是中国散裂中子源工程（CSNS）废束站的重要部件。文中探讨了废束站束窗的材料选择，结构设计和其他相关的问题。设计中采用热分析的方法。通过比较，Glidcop®AL-15 和 316L 被确定为束窗的材料。通过对结构的优化分析，束窗截面被设计为内凸的球面形状，而束窗厚度设定在 1.5~3 毫米之间。经确认，即使在磁铁失效时也能确保束窗安全。分析表明束窗能够很好地满足 CSNS 废束站的使用要求。

**关键词：** 废束站束窗，热分析，束窗截面，束窗厚度，束窗安全